\documentclass[final,5p,times,twocolumn]{elsarticle}

\usepackage{amsmath}
\usepackage{amssymb}
\usepackage{graphicx}
\usepackage{color}
\usepackage{hyperref}
\usepackage{xspace}

\usepackage[mathscr,scaled=1.15]{urwchancal}
\DeclareFontFamily{OT1}{pzc}{}
\DeclareFontShape{OT1}{pzc}{m}{it}%
{<-> s * [1.15] pzcmi7t}{}
\DeclareMathAlphabet{\mathpzc}{OT1}{pzc}{m}{it}

\definecolor{purple}{rgb}{0.5,0,0.5}
\definecolor{blue}{rgb}{0.0,0,0.9}

\biboptions{sort&compress}

\journal{Physics Letters B}

\begin{document}

\begin{frontmatter}

\title{Sketching the pion's valence-quark generalised parton distribution}

\author[CEA]{C. Mezrag}
\author[UA]{L. Chang}
\author[CEA]{H. Moutarde }
\author[ANL]{C.\,D. Roberts}
\author[Huelva]{J. Rodr\'iguez-Quintero}
\author[CEA]{F. Sabati\'e}
\author[JARA]{S.\,M. Schmidt}

\address[CEA]{Centre de Saclay, IRFU/Service de Physique Nucl\'eaire, F-91191 Gif-sur-Yvette, France}
\address[UA]{CSSM, School of Chemistry and Physics
University of Adelaide, Adelaide SA 5005, Australia}
\address[ANL]{Physics Division, Argonne National Laboratory, Argonne, Illinois 60439, USA}
\address[Huelva]{Departamento de F\'isica Aplicada, Facultad de Ciencias Experimentales, Universidad de Huelva, Huelva E-21071, Spain}
\address[JARA]{Institute for Advanced Simulation, Forschungszentrum J\"ulich and JARA, D-52425 J\"ulich, Germany}

\date{18 November 2014}

\begin{abstract}
$\,$\\[-7ex]\hspace*{\fill}{\emph{Preprint nos}. ADP-14-37/T896; IRFU-14-48}\\[1ex]
%
In order to learn effectively from measurements of generalised parton distributions (GPDs), it is desirable to compute them using a framework that can potentially connect empirical information with basic features of the Standard Model.  We sketch an approach to such computations, based upon a rainbow-ladder (RL) truncation of QCD's Dyson-Schwinger equations and exemplified via the pion's valence dressed-quark GPD, $H_\pi^{\rm v}(x,\xi,t)$.  Our analysis focuses primarily on $\xi=0$, although we also capitalise on the symmetry-preserving nature of the RL truncation by connecting $H_\pi^{\rm v}(x,\xi=\pm 1,t)$ with the pion's valence-quark parton distribution amplitude.  We explain that the impulse-approximation used hitherto to define the pion's valence dressed-quark GPD is generally invalid owing to omission of contributions from the gluons which bind dressed-quarks into the pion.  A simple correction enables us to identify a practicable improvement to the approximation for $H_\pi^{\rm v}(x,0,t)$, expressed as the Radon transform of a single amplitude. Therewith we obtain results for $H_\pi^{\rm v}(x,0,t)$ and the associated impact-parameter dependent distribution, $q_\pi^{\rm v}(x,|\vec{b}_\perp|)$, which provide a qualitatively sound picture of the pion's dressed-quark structure at an hadronic scale.  We evolve the distributions to a scale $\zeta=2\,$GeV, so as to facilitate comparisons in future with results from experiment or other nonperturbative methods.
\end{abstract}

\end{frontmatter}


\noindent\textbf{1.$\;$Introduction}.
Quarks were discovered in a series of deep inelastic scattering (DIS) experiments at the Stanford Linear Accelerator Center \cite{Friedman:1991nq,Taylor:1991ew,Kendall:1991np}.  When analysed in the infinite momentum frame; i.e., treating the target as an extremely rapidly moving object, such experiments yield parton distribution functions (PDFs).  PDFs are probability densities, which reveal how partons within the speeding target share the bound-state's gross properties; e.g., there are PDFs that describe the distributions over the target's constituent partons of the total longitudinal momentum and helicity.  Crucially, this probability interpretation is only valid in the infinite-momentum frame owing to its connection with quantisation on the light-front \cite{Keister:1991sb,Coester:1992cg,Brodsky:1997de}, a procedure that ensures, \emph{inter alia}, particle number conservation.

A good deal is known about hadron light-front structure after more than forty years of studying PDFs.  Notwithstanding that, much more needs to be understood, particularly, e.g., in connection with the distribution of helicity \cite{Myhrer:2009uq,Jimenez-Delgado:2013sma}.  Moreover, PDFs only describe hadron light-front structure incompletely because inclusive DIS measurements do not yield information about the distribution of partons in the plane perpendicular to the bound-state's total momentum; i.e., within the light front.  Such information is expressed in generalised parton distributions (GPDs) \cite{Dittes:1988xz,Mueller:1998fv,Radyushkin:1996nd,Ji:1996nm}, which are accessible via deeply virtual Compton scattering on a target hadron, $T$; viz., $\gamma^\ast(q) T(p) \to \gamma^\ast(q^\prime)T(p^\prime)$, so long as at least one of the photons [$\gamma^\ast(q)$, $\gamma^\ast(q^\prime)$] possesses large virtuality, and in the analogous process of deeply virtual meson production: $\gamma^\ast(q) T(p) \to M(q^\prime)T(p^\prime)$.  Importantly [see Sect. 2], GPDs connect PDFs with hadron form factors because any PDF may be recovered as a forward limit of the relevant GPD and any hadron elastic form factors can be expressed via a GPD-based sum rule.  The potential that GPDs hold for providing manifold insights into hadron structure has led to intense experimental and theoretical activity \cite{Burkardt:2002hr,Diehl:2003ny,Belitsky:2005qn,Boffi:2007yc,Guidal:2013rya}.

Most of the constraints that apply to GPDs are fulfilled when the GPD is written as a double distribution
\cite{Mueller:1998fv,Radyushkin:1998es,Radyushkin:1998bz}, which is equivalent to expressing the GPD as a Radon transform \cite{Teryaev:2001qm}:
\begin{equation}
\label{radon}
H(x,\xi,t) = \int_{|\alpha|+|\beta|\leq 1} \rule{-5ex}{0ex} d\alpha \, d\beta \,
\delta(x-\alpha-\xi\beta)\, [F(\alpha,\beta,t) + \xi\, G(\alpha,\beta,t)]\,,
\end{equation}
where the variables $x$, $\xi$, $t$ are defined following Eq.\,\eqref{eq-def-GPD-H-spinless-target} and, at leading-twist, $F$, $G$ have operator definitions analogous to the GPD itself.  In order to obtain insights into the nature of hadron GPDs, it has been common to model the Radon amplitudes, $F$, $G$, following Refs.\,\cite{Musatov:1999xp}.  This approach has achieved some phenomenological success \cite{Guidal:2013rya,Mezrag:2013mya}; but more flexible parametrisations enable a better fit to data \cite{Kumericki:2008di}.  Such fits played a valuable role in establishing the GPD framework;
but if one wishes to use measured GPDs as a means by which to validate our basic perception of strong interactions in the Standard Model, then data fitting is inadequate.  Instead, it is necessary to compute GPDs using a framework that possesses a direct connection with QCD.  This observation is highlighted by experience drawn from the simpler case of the pion's valence-quark PDF \cite{Chang:2014lva}.  Herein, we therefore adopt a different approach, sketching a procedure for the computation of hadron GPDs based on the example provided by the pion's valence-quark PDF.

\smallskip

\noindent\textbf{2.$\;$General features of pion GPDs}.
{F}rom a quark model perspective, in the isospin symmetric limit, the pion is a  quantum mechanical bound-state of two equal-mass constituents and it is therefore the simplest hadronic bound-state.  That is a misapprehension, however.  Owing to the connection between pion properties and dynamical chiral symmetry breaking (DCSB); i.e., its dichotomous nature as a Goldstone mode and relativistic bound-state \cite{Maris:1997hd,Qin:2014vya}, a veracious description of the pion is only possible within a framework that faithfully expresses symmetries and their breaking patterns.  The Dyson-Schwinger equations (DSEs) fulfill this requirement \cite{Chang:2011vu,Bashir:2012fs,Cloet:2013jya} and hence we employ that framework to compute pion properties herein.

Notwithstanding the complex nature of the pion bound-state, it is still a $J=0$ system and hence for a vector probe there is only one GPD associated with a quark $q$ in the pion ($\pi^\pm$, $\pi^0$).  It is defined by the matrix element
\begin{eqnarray}
\label{eq-def-GPD-H-spinless-target}
\nonumber\lefteqn{
H^q_{\pi}( x, \xi, t ) =  \int \frac{\mathrm{d}^4z}{4\pi} \, e^{i x P\cdot z}\,
\delta(n\cdot z) \, \delta^2(z_\perp)}\\
&& \quad \times \,\langle\pi(P_+)| \bar{q}\left(-z/2\right)n\cdot \gamma \; 
q\left(z/2\right) |\pi(P_-)\rangle,
\end{eqnarray}
where: $k$, $n$ are light-like four-vectors, satisfying $k^2=0=n^2$, $k\cdot n=1$; $z_\perp$ represents that two-component part of $z$ annihilated by both $k$, $n$; and $P_\pm = P \pm \Delta/2$.
%
In Eq.\,\eqref{eq-def-GPD-H-spinless-target}, $\xi = -n\cdot \Delta/[2 n\cdot P]$ is the ``skewness'', $t=-\Delta^2$ is the momentum transfer, and $P^2 = t/4-m_\pi^2$, $P\cdot \Delta=0$.
The GPD also depends on the resolving scale, $\zeta$.  Within the domain upon which perturbation theory is valid, evolution to another scale $\zeta^\prime$ is described by the ERBL equations \cite{Efremov:1979qk,Lepage:1980fj} for $|x|<\xi$ and the DGLAP equations \cite{Dokshitzer:1977,Gribov:1972,Lipatov:1974qm,Altarelli:1977} for $|x|>\xi$, where $\xi \geq 0$.

In order to produce quantities that are gauge invariant for all values of $z$, Eq.\,\eqref{eq-def-GPD-H-spinless-target} should contain a Wilson line, ${\cal W}[-z/2,z/2]$, between the quark fields.  Notably, for any light-front trajectory, ${\cal W}[-z/2,z/2]\equiv 1$ in lightcone gauge: $n\cdot A=0$, and hence the Wilson line does not contribute in this case.  On the other hand, light-cone gauge is seldom practicable in either model calculations or quantitative nonperturbative analyses in continuum QCD.  Indeed, herein, as typical of nonperturbative DSE studies, we employ Landau gauge because, \emph{inter alia} \cite{Bashir:2008fk,Bashir:2009fv}: it is a fixed point of the renormalisation group; and a covariant gauge, which is readily implemented in numerical simulations of lattice-QCD.  It is therefore significant that ${\cal W}[-z/2,z/2]$ is not quantitatively important in the calculation of the leading-twist contributions to numerous matrix elements \cite{Kopeliovich:2011rv}.

It is worth recapitulating here upon some general properties of GPDs.  Most generally, Poincar\'e covariance entails that GPDs are only nonzero on $x\in (-1,1)$.  Moreover, owing to time-reversal invariance, $H^{q}(x,\xi,t) =H^{q}(x,-\xi,t)$.  Kinematically, the skewness is bounded: $\xi\in[-1,1]$, but $\xi\in[0,1]$ for all known processes that provide empirical access to GPDs.

Focusing on the pion, a charge conjugation mapping between charged states entails $H^{u,d}_{\pi^+}(x,\xi,t) =  - H^{u,d}_{\pi^-}(-x,\xi,t)$; and consequently, in the isospin symmetric limit:
\begin{equation}
H^u_{\pi^+}(x,\xi,t) = - H^d_{\pi^+}(-x,\xi,t).
\label{eq:G-parity-GPD}
\end{equation}
It follows that the isospin projections:
\begin{equation}
H^{I}(x,\xi,t):=
H_{\pi^+}^u(x,\xi,t) + (-1)^I H_{\pi^+}^d(x,\xi,t), I=0,1\,, \label{eq:H_isoscalar_both}
\end{equation}
have well-defined symmetry properties under $x\leftrightarrow -x$; viz.,
$H^{0}$ is odd and $H^{1}$ is even.

Returning to the definition in Eq.\,\eqref{eq-def-GPD-H-spinless-target}, it is plain that if one considers the forward limit: $\xi=0$, $t=0$, then $x$ is Bjorken-$x$ and the GPD reduces to a PDF; viz.,
\begin{equation}
H_\pi^q(x,0,0) = \left\{
\begin{array}{rr}
q^\pi(x), & x>0\\
-\bar q^\pi(-x), & x<0
\end{array}\right.\,.
\label{PDFconnection}
\end{equation}
Moreover, irrespective of the value of $\xi$, the electromagnetic pion form factor may be computed as
\begin{align}
F_{\pi^+}(\Delta^2) &= \int_{-1}^1 dx\, [ e_u  H_{\pi^+}^u(x,\xi,-\Delta^2) + e_d  H_{\pi^+}^d(x,\xi,-\Delta^2) ]\quad\label{Fpiconnection1}\\
&=: e_u F_{\pi^+}^u(\Delta^2) + e_d F_{\pi^+}^d(\Delta^2)
= F_{\pi^+}^u(\Delta^2)\,, \label{Fpiconnection2}
\end{align}
where $e_{u,d}$ are the quark electric charges in units of the positron charge and we have used Eq.\,\eqref{eq:G-parity-GPD} to show $F_{\pi^+}^d(\Delta^2) = -  F_{\pi^+}^u(\Delta^2)$.  Additional information may be found elsewhere \cite{Mezrag:2014tva}.

\smallskip

\noindent\textbf{3.$\;$Heuristic Example}.
Imagine a bound-state of two scalar particles with effective mass $\sigma$ and suppose that the interaction between them is such that it produces a light-front wave function of the form ($\bar x=1-x$):
\begin{equation}
\label{psimodel}
\psi(x,k_\perp^2) = \sqrt{\frac{15}{2\pi \,\sigma^2 }}\,\frac{\sqrt{x \bar x}}{1+k_\perp^2/(4 \,\sigma^2 x\bar x)}\theta(x)\theta(\bar x)\,.
\end{equation}
(A merit of considering a bound-state of scalar constituents is that in describing the wave function of the composite system one avoids the complication of Melosh rotations, which arise in treating spin states in light-front quantum mechanics \cite{Coester:1992cg}.)  If the skewness is zero, in which case the momentum transfer is purely light-front transverse, then the GPD for this system can be written as a wave function overlap \cite{Burkardt:2000za,Diehl:2000xz,Burkardt:2002hr,Diehl:2003ny}:
\begin{equation}
\label{overlapGPD}
H_\sigma(x,0,-\Delta_\perp^2) = \int d^2 k_\perp \, \psi(x,k_\perp+(1-x) \Delta_\perp) \, \psi(x,k_\perp)\,.
\end{equation}
This entails
\begin{equation}
\label{HPositive}
\{ H_\sigma(x,0,\Delta_\perp^2)>0 : x\in[-1,1], \Delta_\perp^2\geq 0 \}\,.
\end{equation}

Owing to the simplicity of the starting point, Eqs.\,\eqref{psimodel} and \eqref{overlapGPD} allow one to obtain an algebraic expression for the GPD; viz., with $z^2 =\Delta_\perp^2 (1-x)/4x\sigma^2 $, then
\begin{align}
&H_\sigma(x,0,-\Delta_\perp^2)  = 30 (1-x)^2 x^2
{\cal C}(z)\,\theta(x)\theta(\bar x)\,,
\\
&{\cal C}(z) = \frac{\ln\left[ \left(z^3+\left(z^2+1\right)
   \sqrt{z^2+4}+3 z\right)/
   \left(\sqrt{z^2+4}-z\right)\right]}{z \sqrt{z^2+4}}\,.
   \label{eqCz}
\end{align}
Some further analysis reveals that $C(z)$ decreases monotonically away from its maximum value ${\cal C}(z=0)=1$.  Consequently, $H_\sigma(x,0,0)= 30 (1-x)^2 x^2\theta(x)\theta(\bar x) $, which is an excellent approximation to the pion's valence dressed-quark PDF \cite{Chang:2014lva}; and whilst the maximum of $H_\sigma(x,0,-\Delta^2)$ lies at $x=1/2$ for $\Delta^2=0$, this peak shifts to $x=1$ with increasing $\Delta^2$, consistent with an expectation that for $\Delta^2\gg \sigma^2$ the interaction probability is largest when the probe and hadron are collinear \cite{Burkardt:2002hr}.

The Hankel transform:
\begin{equation}
\label{defqxb}
q_\sigma(x,|\vec{b}_\perp|) =
\int \frac{d|\Delta_\perp|}{2\pi}|\Delta_\perp| J_0(|\vec{b}_\perp||\vec{\Delta}_\perp|) H(x,0,-\Delta_\perp^2)\,,
\end{equation}
defines the system's impact-parameter-dependent (IPD) GPD \cite{Burkardt:2000za}. It is a density that describes the probability of finding a parton within the light-front at a transverse position $\vec{b}_\perp$ from the hadron's centre of transverse momentum (CoTM).  Since $H(x,0,-\Delta_\perp^2)$ is a positive-definite, monotonically decreasing function of $\Delta_\perp^2$ for each $x$, the global maximum of $q_\sigma(x,|\vec{b}_\perp|)$ is located at $|\vec{b}_\perp|=0$ and $q_\sigma(x,|\vec{b}_\perp|)$ is a monotonically decreasing, positive-definite function away from that maximum.

The value of $x$ at which the global maximum in $q_\sigma(x,|\vec{b}_\perp|)$ occurs is determined by the system's dynamics.  Considering the hadron's valence dressed-parton structure, one extreme is achieved if ${\cal C}(z)$ is independent of $x$: the maximum of $q_\sigma(x,|\vec{b}_\perp|)$ is then located at $(x=1/2,|\vec{b}_\perp|=0)$.  In realistic cases, the necessary $(x,\Delta_\perp^2)$ correlations in ${\cal C}(z)$ and this function's general properties act to shift the maximum to $x>1/2$.  Using Eqs.\,\eqref{psimodel} and \eqref{overlapGPD}, $q_\sigma(x,|\vec{b}_\perp|)$ peaks at $(x=0.72,|\vec{b}_\perp|=0)$.  One may also consider the path followed by the maximum as one increases $|\vec{b}_\perp|$ away from zero.  To that end, observe from Eq.\,\eqref{defqxb} that for $|\vec{b}_\perp|\gg 1/\sigma$ the $x$-dependence of $q_\sigma(x,|\vec{b}_\perp|)$ is dominated by $H(x,0,-\Delta_\perp^2 \simeq 0)$, which peaks at $x=1/2$.  The nature of ${\cal C}(z)$ then entails that the peak in the valence dressed-quark IPD GPD drifts monotonically toward $x=1/2$ as $\sigma |\vec{b}_\perp|\to 0$.

\smallskip

\noindent\textbf{4.$\;$Pion's valence dressed-quark GPD}.
In order to compute an approximation to the valence-quark piece of the GPD expressed in Eq.\,\eqref{eq-def-GPD-H-spinless-target} we adapt the method used successfully elsewhere to compute the pion's valence-quark distribution function \cite{Chang:2014lva} and elastic form factor \cite{Chang:2013nia}.  Consider, therefore,
\begin{align}
\nonumber
2 H_{\pi}^{\rm v}(x,\xi,t)& =
 N_c \mathrm{tr}\rule{-0.5ex}{0ex}
 \int_{d\ell}\,\delta_n^{xP}(\ell)\,i{\Gamma}_\pi(\ell_+^{\rm R};-P_+ )\,\\
&
\times
S(\ell_+) \,in\cdot\Gamma(\ell_+,\ell_-) \, S(\ell_-) i\Gamma_\pi(\ell_-^{\rm R}; P_- )\,,
\label{eq:TriangleDiagrams}
\end{align}
where $\int_{d\ell} := \int \frac{d^4\ell}{(2\pi)^4}$ is a translationally invariant regularisation of the integral; $\delta_n^{xP}(\ell):= \delta(n\cdot \ell - x n\cdot P)$; the trace is over spinor indices; $\eta\in[0,1]$, $\bar\eta=1-\eta$; $\ell_+^{\rm R}=\bar\eta\ell_+ +\eta\ell_P$,
$\ell_-^{\rm R}=\eta\ell_- +\bar\eta\ell_P$,
$\ell_\pm = \ell \pm \Delta/2$, $\ell_P=\ell-P$.  (N.B.\ Owing to Poincar\'e covariance, no observable can legitimately depend on $\eta$; i.e., the definition of the relative momentum.)
So long as each of the dressed-quark propagators, $S(\ell)$, on the right-hand-side (rhs) of Eq.\,\eqref{eq:TriangleDiagrams} is computed using the rainbow truncation of QCD's gap equation, and both the pion Bethe-Salpeter amplitudes, $\Gamma_\pi(\ell;P)$, and the dressed-quark-photon vertex, $\Gamma_\mu(\ell_f,\ell_i)$, are calculated in the associated ladder truncation of the relevant Bethe-Salpeter equations then $H_{\pi}(x,\xi,t)$, thus computed and inserted in Eq.\,\eqref{Fpiconnection1}, provides the leading-order contribution to the pion's electromagnetic form factor in the most widely used, symmetry preserving truncation of QCD's DSEs: the rainbow-ladder (RL) truncation \cite{Munczek:1994zz,Bender:1996bb}, whose strengths and limitations are detailed elsewhere \cite{Chang:2011vu,Bashir:2012fs,Cloet:2013jya}.

Given its connection with a reliable scheme for computing $F_\pi(Q^2)$, it was long thought \cite{Hecht:2000xa,Nguyen:2011jy} that Eq.\,\eqref{eq:TriangleDiagrams} would also be an adequate starting point for computation of the pion's valence-quark PDF, $q_V^\pi(x)$.
However, as explained in Ref.\,\cite{Chang:2014lva}, that is not true: Eq.\,\eqref{eq:TriangleDiagrams} derives from the \emph{handbag} \emph{diagram} contribution to $q_V^\pi(x)$ and that impulse approximation is incomplete because it omits a fraction of the contributions from gluons which bind dressed-quarks into the pion.

Since Eq.\,\eqref{eq:TriangleDiagrams} is incomplete for $q_V^\pi(x)$ it is necessarily also inadequate for computation of $H_\pi^{\rm v}(x,\xi,t)$.  Importantly, we have found that the flaw is expressed more forcefully as ${\mathpzc t}:=\Delta_\perp^2$ grows: one can obtain $H_\pi^{\rm v}(x,0,-{\mathpzc t})<0$, which is physically impossible, as explained in connection with Eqs.\,\eqref{overlapGPD}, \eqref{HPositive}.  The precise form for the correction to $q_V^\pi(x)$ is known but the related amendment to Eq.\,\eqref{eq:TriangleDiagrams} is still being sought [Eq.\,\eqref{HCorrection} below is a rudimentary model].  We will therefore focus primarily on $\xi=0$, be guided by Eq.\,\eqref{eq:TriangleDiagrams}, and mention and ameliorate its failings where appropriate, drawing on the insights gained from the example in Sect.\,3.  Notably, the defects of Eq.\,\eqref{eq:TriangleDiagrams} are typically overlooked in extant continuum computations of the pion's GPD \cite{Tiburzi:2002tq,Broniowski:2003rp,Ji:2006ea,Broniowski:2007si,Frederico:2009fk}): Refs.\,\cite{Broniowski:2003rp,Broniowski:2007si} deliver a form for $q(x,|\vec{b}_\perp|)$ that is not positive definite.

In order to gain novel insights into pion structure, we use the following algebraic forms for the dressed-quark and pion elements in Eq.\,\eqref{eq:TriangleDiagrams} $[\Delta_M(\ell^2)=1/(\ell^2+M^2)]$ \cite{Chang:2013pq}:
\begin{subequations}
\label{NakanishiASY}
\begin{eqnarray}
\label{eq:sim1}
S(\ell) & = &[-i\gamma\cdot \ell+M]\Delta_M(\ell^2)\,,\\
\rho_\nu(z) & = & \frac{1}{\sqrt{\pi}}\frac{\Gamma(v+3/2)}{\Gamma(\nu+1)}(1-z^2)^\nu\,,\\
\label{eq:sim2}
%
%
\mathpzc{n}_\pi \Gamma_\pi(\ell^{\rm R}_\pm;\pm P) & = & i\gamma_5\int^1_{-1}dz\, \rho_\nu(z) \, \hat\Delta^\nu_M(\ell^2_{z\pm})\,,
\label{NoF}
\end{eqnarray}
\end{subequations}
where $M$ is a dressed-quark mass-scale; $\hat\Delta_M(\ell^2) = M^2 \Delta_M(\ell^2)$; $\ell_{z\pm}=\ell^{\rm R}_\pm + (z \pm 1) P/2$ and we work in the chiral limit ($P^2=0=\hat m$, where $\hat m$ is the current-quark mass); and $\mathpzc{n}_\pi$ is the Bethe-Salpeter amplitude's canonical normalisation constant.  Owing to the simplicity of Eqs.\,\eqref{NakanishiASY}, one can reasonably employ $\Gamma_\mu(\ell_+,\ell_-) = \gamma_\mu P_T({\mathpzc t} = \Delta_\perp^2)$, where $P_T({\mathpzc t})$ is the vertex dressing function described in Eqs.\,(24)--(28) of Ref.\,\cite{Roberts:2011wy}.

Working with the input specified in connection with Eqs.\,\eqref{NakanishiASY}, we computed the triangle diagram result for $H_\pi^{\rm v}(x,\xi,t)$.  As detailed elsewhere \cite{Mezrag:2014tva}, that task was completed by deriving an expression for the Mellin moments of $H_\pi^{\rm v}(x,\xi,t)$ from Eq.\,\eqref{eq:TriangleDiagrams} and introducing five Feynman parameters ($x, y, u, v, w$), defined in the domain $[0, 1]$, and two convolution parameters $z, z^\prime\in [-1, +1]$, so that the momentum integrals could be computed analytically.  Inspecting the result, one can then determine Radon amplitudes for use in Eq.\,\eqref{radon} that are consistent with those moments: the amplitudes vanish outside $\Omega=\{(\alpha,\beta): |\alpha| + |\beta|\leq 1\}$, $F(\alpha,\beta)$ is an even function of $\beta$ and $G(\alpha,\beta)$ is odd.  Eq.\,\eqref{radon} then entails that $H_\pi^{\rm v}(x,\xi,t)$ complies with the known constraints on polynomiality in $\xi$, vanishes outside $x\in[-\xi,1]$ and is continuous at $x=\xi$.

We note now that when considering the comprehensive GPD defined by Eq.\,\eqref{eq-def-GPD-H-spinless-target}, one may write with complete generality:
\begin{subequations}
\label{Hgeneral}
\begin{align}
H_\pi(x,0,-{\mathpzc t}) &= H_\pi(x,0,0) {\cal N}({\mathpzc t})
{\cal C}_\pi(x,{\mathpzc t}) F_\pi({\mathpzc t}) \,, \\
1 & =  {\cal N}({\mathpzc t}) \int_{-1}^1dx\, H_\pi(x,0,0)
{\cal C}_\pi(x,{\mathpzc t})\,,
\end{align}
\end{subequations}
so that all $(x,{\mathpzc t})$ correlations in $H_\pi$ are expressed by ${\cal C}_\pi(x,{\mathpzc t})$, which is necessarily non-unity in any physical system \cite{Burkardt:2002hr}.  It is plain from Eq.\,\eqref{radon} that only $F(\alpha,\beta,t)$ contributes when $\xi=0$.

\begin{figure}[t]

\centerline{\includegraphics[width=0.7\linewidth]{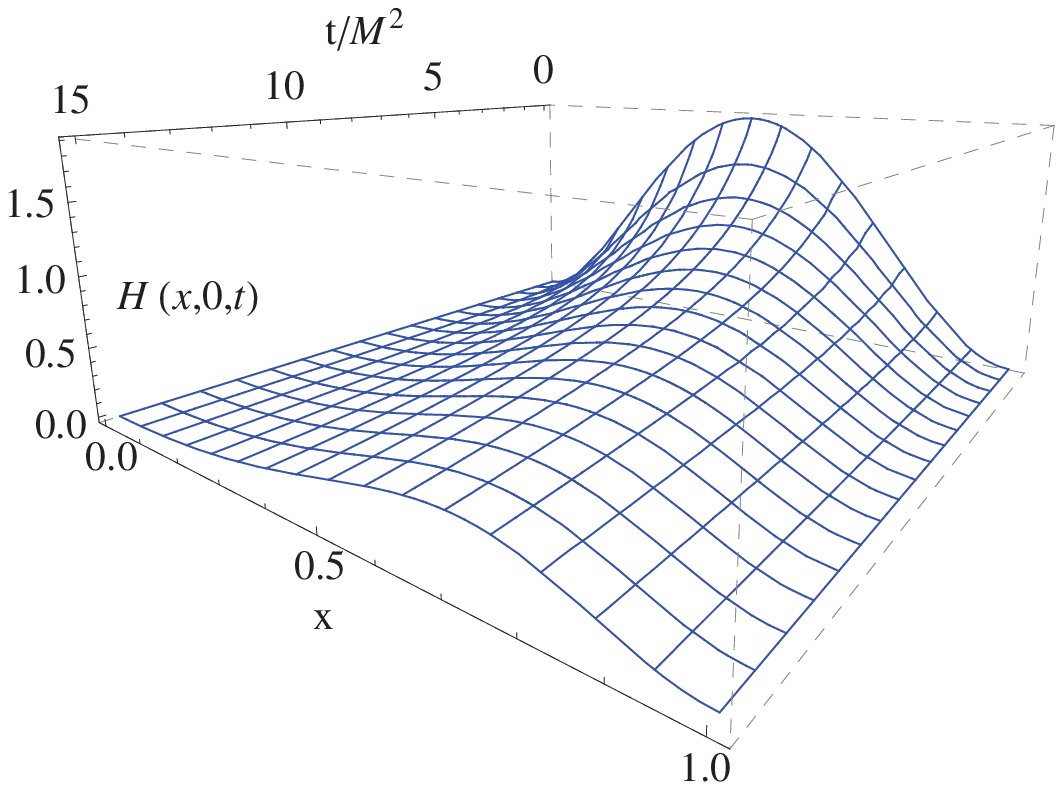}}
\centerline{\includegraphics[width=0.7\linewidth]{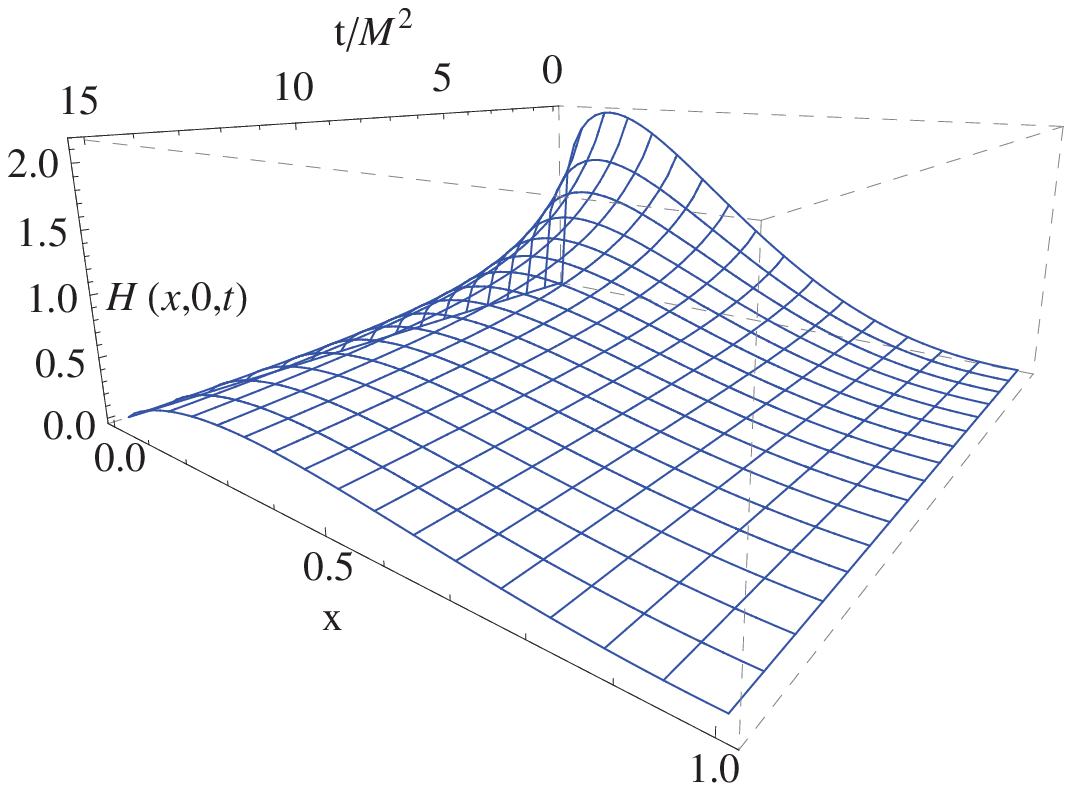}}

\caption{Pion valence dressed-quark GPD, $H_\pi^{\rm v}(x,0,-{\mathpzc t})$, defined by Eqs.\,\eqref{Hgeneral}, \eqref{Cfinal}.  \emph{Upper panel} -- result obtained at the model scale, $\zeta_H=0.51\,$GeV; and \emph{lower panel} -- GPD evolved to $\zeta_2=2\,$GeV using leading-order DGLAP equations, see Sect.\,5.
\label{figHx0tz0z2}}
\end{figure}

In order to continue, we augment Eq.\,\eqref{eq:TriangleDiagrams} by
[$d_\pm^n = n\cdot \partial_{\ell_\pm^{\rm R}}$]
\begin{align}
\nonumber  H_\pi^{\rm C}&(x,0,-{\mathpzc t}) =
\frac{1}{2} N_c {\rm tr}\!\!\!
\int_{d\ell}\,\delta_n^{xP}(\ell)\left[
d_+^n\Gamma_\pi(\ell_+^{\rm R};-P_+)
S(\ell_P)\Gamma_\pi(\ell_-^{\rm R};P_-) \right.\\
&
\left.  \times S(\ell_-) + \Gamma_\pi(\ell_+^{\rm R};-P_+)S(\ell_P)
d_-^n\Gamma_\pi(\ell_-^{\rm R};P_-) S(\ell_-)
\right]\,. \label{HCorrection}
\end{align}
This \emph{Ansatz} extends the handbag diagram correction for $q_V^\pi(x)$ identified in Ref.\,\cite{Chang:2014lva} to ${\mathpzc t}>0$; and, in connection with the valence dressed-quark GPD, it can be expressed via a Radon amplitude $F^{\rm C}(\alpha,\beta,t)$ which preserves the good features of the kindred amplitude produced by Eq.\,\eqref{eq:TriangleDiagrams}.  Summing the contributions from Eqs.\,\eqref{eq:TriangleDiagrams}, \eqref{HCorrection}, the net result has the form
\begin{subequations}
\begin{align}
\label{Fdefine}
F(\alpha,\beta,-{\mathpzc t})
&=\phi(\alpha,\beta,{\mathpzc t})^2[F_{\rm S}(\alpha,\beta)
+ {\mathpzc t} V(\alpha,\beta)\phi(\alpha,\beta,{\mathpzc t})] ,\\
\phi(\alpha,\beta,{\mathpzc t}) & =
1/[1+(t/[4M^2])(1-\alpha+\beta)(1-\alpha-\beta)]\,,
\end{align}
\end{subequations}
where the $F_{\rm S}$ component yields $H_\pi^{\rm v}(x,0,0)=q_V^\pi(x)$ in Ref.\,\cite{Chang:2014lva} and that with $V$ is responsible for all violations of Eq.\,\eqref{HPositive}.

\begin{figure}[t]

\centerline{\includegraphics[width=0.7\linewidth]{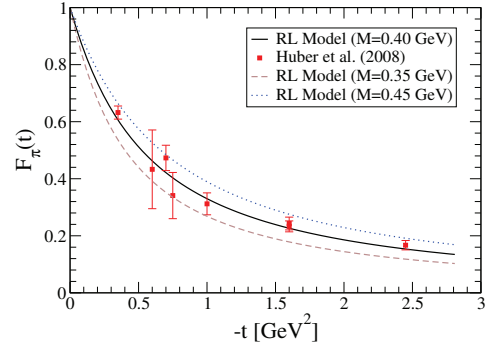}}

\caption{Pion electromagnetic form factor obtained from $H_\pi^{\rm v}(x,0,-{\mathpzc t})$, defined by Eqs.\,\eqref{Hgeneral}, \eqref{Cfinal}, which is deliberately consistent with the result determined using Eqs.\,\eqref{eq:TriangleDiagrams}, \eqref{NakanishiASY} and associated definitions.  The data are described in Ref.\,\protect\cite{Huber:2008id}. The most favourable comparison is obtained with $M=0.40\,$GeV in Eqs.\,\eqref{NakanishiASY} and the band shows results with $M=0.40\pm 0.05\,$GeV.
\label{figFpit}}
\end{figure}

Acting upon these observations, we define an ameliorated RL approximation to the pion's valence dressed-quark GPD as the function obtained by: setting $V\equiv 0$ in Eq.\,\eqref{Fdefine}; and, for added simplicity, working with $\phi(\alpha,\beta=0,{\mathpzc t})$ whilst keeping the form of $F_\pi({\mathpzc t})$ computed directly from Eq.\,\eqref{eq:TriangleDiagrams}.  Namely, via Eq.\,\eqref{radon}, our valence-quark GPD is given by Eq.\,\eqref{Hgeneral} with
\begin{equation}
\label{Cfinal}
{\cal C}(x,{\mathpzc t}) = 1/[ 1 +({\mathpzc t}/[4M^2]) (1-x)^2 ]^2 .
\end{equation}

Our computed GPD is depicted in the upper panel of Fig.\ref{figHx0tz0z2}.  Notably, the properties described in association with Eqs.\,\eqref{HPositive}--\eqref{eqCz} are evident, and this GPD naturally reproduces the pion valence dressed-quark distribution function obtained in Ref.\,\cite{Chang:2014lva}.

The pion form factor associated with our GPD is drawn in Fig.\,\ref{figFpit}.  A fit to the result is provided by
\begin{equation}
\label{eqFpit}
F_\pi({\mathpzc t}=M^2 z)= \frac{1+0.16 z}{1+0.44 z + 0.060 z^2 + 0.00033 z^3}\,.
\end{equation}
At large-$\mathpzc t$ it behaves as $1/{\mathpzc t}^2$, whereas the correct power-law dependence is $1/{\mathpzc t}$ \cite{Farrar:1979aw,Efremov:1979qk,Lepage:1980fj}.  The power-law is wrong because Eq.\eqref{NoF} omits terms that have been described as representing the pion's pseudovector components \cite{Maris:1998hc}, which are necessarily nonzero in a complete picture of the physical pion \cite{Maris:1997hd,Qin:2014vya}.  Notwithstanding that, it is valid and useful to compare the prediction with contemporary data and thereby determine a sensible value for our model's dressed-quark mass-scale: the best comparison is obtained with $M=0.4\,$GeV.  Notably, this scale is typical of the dressed-quark mass function in QCD \cite{Bhagwat:2003vw,Bowman:2005vx,Bhagwat:2006tu}.

\begin{figure}[t]

\centerline{\includegraphics[width=0.7\linewidth]{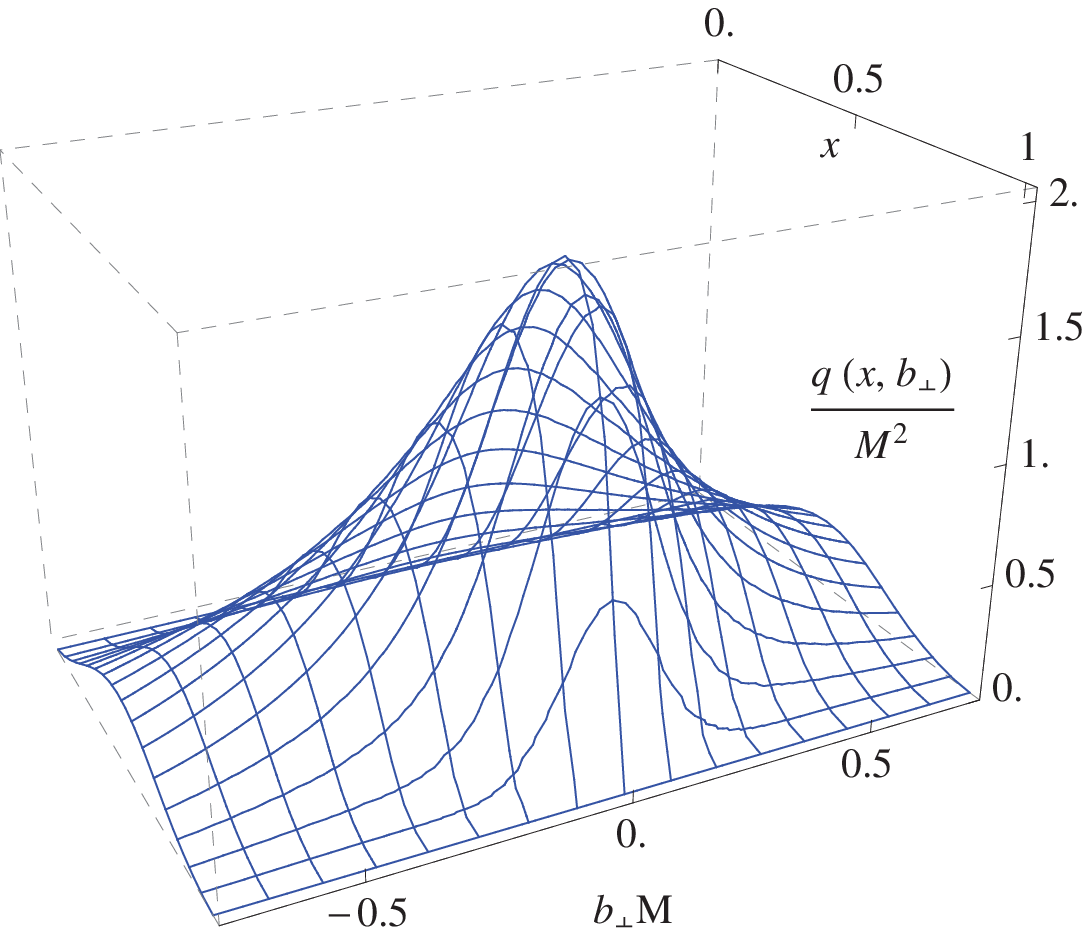}}\vspace*{-2ex}

\centerline{\includegraphics[width=0.7\linewidth]{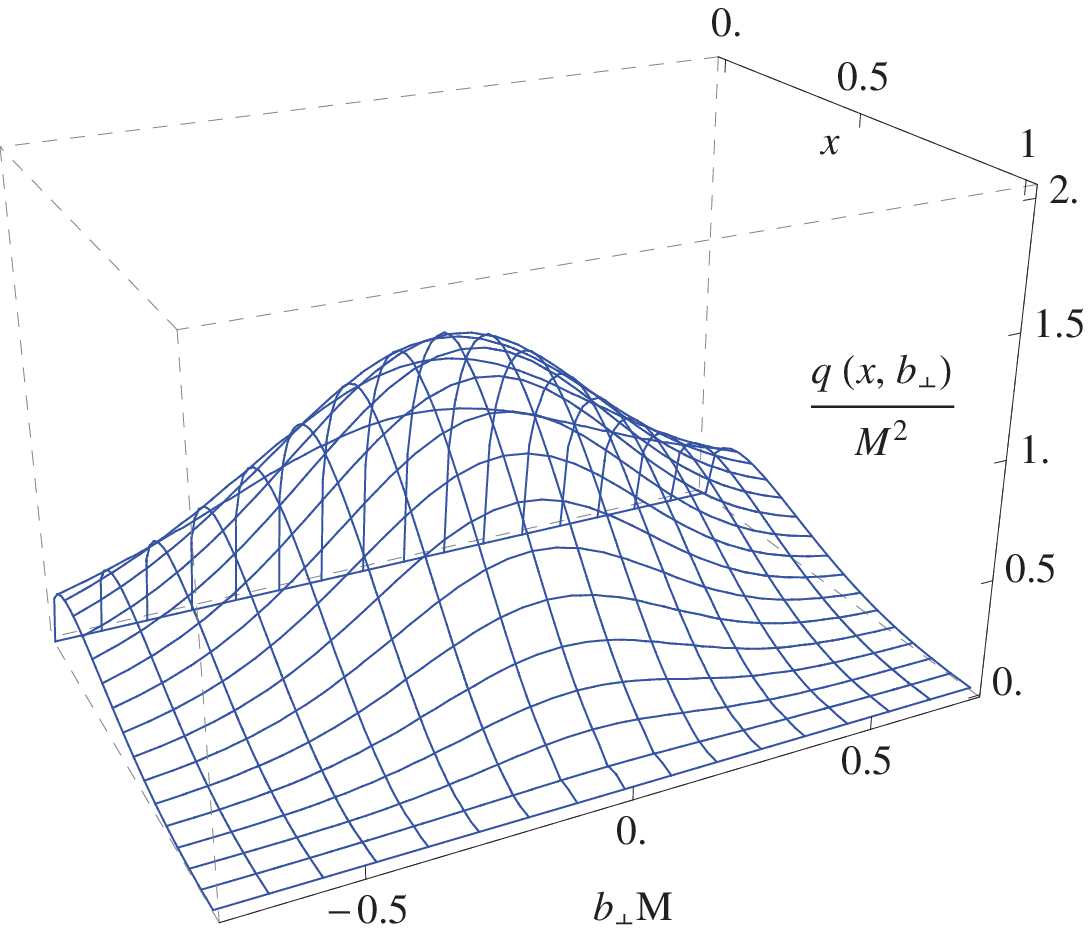}}

\caption{Pion's valence dressed-quark GPD in impact parameter space, $q_\pi^{\rm v}(x,|\vec{b}_\perp|;\zeta)$.  \emph{Upper panel} -- result obtained from $H_\pi^{\rm v}(x,0,-{\mathpzc t};\zeta_H)$ in the top panel of Fig.\,\ref{figHx0tz0z2} using Eq.\,\eqref{defqxb}; and \emph{lower panel} -- analogous result associated with $H_\pi^{\rm v}(x,0,-{\mathpzc t};\zeta_2)$ in the bottom panel of Fig.\,\ref{figHx0tz0z2}, see Sect.\,5.  [N.B.\,$1/M \approx 0.5\,$fm, so $ b_\perp M =0.5$ corresponds to $b_\perp \approx 0.25\,$fm and $ q_\pi^{\rm v}(x,|\vec{b}_\perp|;\zeta)/M^2 =1$ means $q_\pi^{\rm v}(x,|\vec{b}_\perp|;\zeta)\approx 4\,$fm$^{-2}$.]
\label{figqbx2}}
\end{figure}

The IPD GPD derived from $H_\pi^{\rm v}(x,0,-{\mathpzc t})$ in the upper panel of Fig.\,\ref{figHx0tz0z2} is depicted in the upper panel of Fig.\,\ref{figqbx2}.  The global maximum in this valence distribution is located at $(x=0.76,|\vec{b}_\perp|=0)$ and, plainly, the probability to find a dressed-quark is strongly localised around this maximum.  Naturally, for this valence dressed-quark distribution ($d^2 |\vec{b}_\perp| = 2\pi \,d |\vec{b}_\perp|\, |\vec{b}_\perp|$):
\begin{equation}
\begin{array}{c}
\int_{-1}^1dx\!\int_0^\infty d^2 |\vec{b}_\perp|\,x \,q_\pi^{\rm v}(x,|\vec{b}_\perp|)
=\frac{1}{2}.
\end{array}
\end{equation}

\smallskip

\noindent\textbf{5.$\;$Evolution of the GPD}.
As explained elsewhere \cite{Chang:2014lva}, our framework yields a valence-quark GPD that may be associated with an hadronic scale $\zeta_{\rm H}=0.51\,$GeV.  It is worth outlining how the features of this distribution evolve to higher scales.  Given that we have used $\xi=0$, that is readily accomplished by using the DGLAP evolution equations to determine the evolved $x$-profile at each value of ${\mathpzc t}$.  Our aim is to provide a qualitative illustration so, unlike Ref.\,\cite{Chang:2014lva}, we do not augment the valence distribution via the inclusion of gluon or sea-quark contributions.  If desired, one could mask the impact of this omission by focusing on the behaviour of $x H_\pi^{\rm v}(x,0,t)$ and $xq_\pi^{\rm v}(x,|\vec{b}_\perp|)$.

Beginning with the valence dressed-quark GPD in the upper panel of Fig.\,\ref{figHx0tz0z2}, we employed leading-order evolution to obtain $H_\pi^{\rm v}(x,0,-{\mathpzc t};\zeta_2=2\,{\rm GeV})$.\footnote{Any significant differences generated by next-to-leading-order evolution are masked by a 25\% increase in $\zeta_{\rm H}$ \cite{Gluck:1999xe} and hence are immaterial herein.}  The result is depicted in the lower panel of Fig.\,\ref{figHx0tz0z2}.  Evidently, evolution, which adds glue and sea-quarks to the system by exposing this substructure within the dressed-quark, sharpens the peak associated with the global maximum at ${\mathpzc t}=0$ and shifts its location toward $x=0$.  The maximum value at each ${\mathpzc t}\neq 0$ is also shifted toward $x=0$; but outside a neighbourhood of ${\mathpzc t}=0$ the profile in $x$ is progressively flattened with increasing ${\mathpzc t}$.  Notwithstanding this, at any finite $\zeta>\zeta_2$, there will be a ${\mathpzc t}_1$ such that $\forall {\mathpzc t}> {\mathpzc t}_1$ a peak, albeit with much suppressed height, may be said to exist in the neighbourhood $x\simeq 1$: ${\mathpzc t}_1$ increases with $\zeta$.

The last observation leads us to consider the conformal limit of QCD, which is recovered on $\tau \simeq 0$, $\tau=\Lambda_{\rm QCD}/\zeta$.  Within this domain, the valence dressed-quark GPD is $H_\pi^{\rm v}(x,0,0;\tau \simeq 0)=\delta(x)$ \cite{Georgi:1951sr,Gross:1974cs,Politzer:1974fr}.  [Fig.\,\ref{figHx0tz0z2} highlights that this limit is reached slowly because evolution is logarithmic in QCD.]  Eqs.\,\eqref{Hgeneral} then entail:
\begin{equation}
\label{Hvtau0}
H_\pi^{\rm v}(x,0,-{\mathpzc t};\tau \simeq 0) = \delta(x) F_{\pi}({\mathpzc t})\,.
\end{equation}
This is a feature of our approximation to the pion's valence dressed-quark GPD.  It is not a property of the pion's complete GPD, $H_\pi(x,0,t;\tau \simeq 0)$, because the valence GPD is a negligible piece of the full GPD on $\tau\simeq 0$.  That may be seen, e.g., by noting that valence-quarks carry none of the pion's momentum within the conformal domain and hence it is invalid therein to represent $F_\pi(t)$ by an impulse (rainbow-ladder) approximation expressed through the triangle diagram of Eq.\,\eqref{eq:TriangleDiagrams}.

Having determined $H_\pi^{\rm v}(x,0,-{\mathpzc t};\zeta_2)$, it is straightforward to obtain $q_\pi^{\rm v}(x,|\vec{b}_\perp|;\zeta_2)$ from Eq.\,\eqref{defqxb}.  The result is depicted in the lower panel of Fig.\,\ref{figqbx2}: apparently, the maximum is shifted toward $x=0$ and compressed in that direction, the peak height is diminished, and the width of the distribution in $|\vec{b}_\perp|$ is increased.

Each one of these evolution-induced changes may be intuitively understood by reasoning as follows.
First consider a limiting case of an active parton with $x\approx 1$.  This parton carries (almost) all the longitudinal momentum of the hadron.  It therefore \emph{defines} the CoTM and hence cannot be far removed from that centre.  The distribution associated with an $x\approx 1$ parton must therefore be tightly localised around $|\vec{b}_\perp|=0$.
On the other hand, consider the case of an active parton with $x$ reduced toward the location of the global maximum.  The remaining partons within the hadron share in defining the CoTM and hence the active parton is not constrained to lie at $|\vec{b}_\perp|=0$.  Plainly, as a parton's value of $x$ diminishes toward the favoured value, it plays less of a role in determining the CoTM and may therefore possess even larger values of $|\vec{b}_\perp|$.

In the current context, recall that evolution exposes the glue and sea-quark content of a dressed-quark: its identity comes to be shared amongst a host of partons, so that the probability of any one parton carrying $x\approx 1$ is much diminished.  It follows that the global maximum in $q(x,|\vec{b}_\perp|;\zeta)$ must move toward $x=0$ with increasing $\zeta$ and, simultaneously, that the distribution is broadened in $|\vec{b}_\perp|$ on the remaining domain of material support.

\begin{figure}[t]

\centerline{\includegraphics[width=0.7\linewidth]{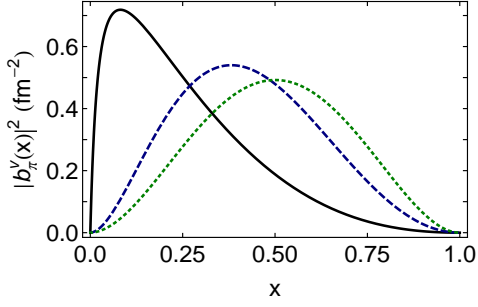}}

\caption{Distribution of pion's mean-square transverse extent, Eq.\,\eqref{bperp2}: (black) solid curve $\langle|\vec{b}_\perp(x;\zeta_2)|^2 \rangle$; and (blue) dashed curve -- $\langle|\vec{b}_\perp(x;\zeta_H)|^2 \rangle$.  The (green) dotted curve is the result obtained at $\zeta_H$ when the correlation function ${\cal C}_\pi(x,t)$ in Eq.\,\eqref{Cfinal} is neglected: comparison with the dashed curve shows that the product \emph{Ansatz} $H_\pi^{\rm v}(x,0,-{\mathpzc t}) = q_{\rm V}^\pi(x) F_\pi({\mathpzc t})$ is generally unreliable.
\label{b2x}}
\end{figure}

The latter effect is illustrated in Fig.\,\ref{b2x}, which depicts
\begin{equation}
\label{bperp2}
\begin{array}{c}
\langle |\vec{b}_\perp(x;\zeta)|^2 \rangle = \int_0^\infty d^2 |\vec{b}_\perp|
\,  q(x,|\vec{b}_\perp|;\zeta)\,|\vec{b}_\perp|^2\,;
\end{array}
\end{equation}
i.e., the $x$-distribution of the pion's mean-square transverse extent: under evolution, the transverse extent narrows at large-$x$ and broadens at small-$x$.  A little consideration reveals that the measure of the curves in Fig.\,\ref{b2x} is independent of the scale $\zeta$ because evolution is an operation that preserves the area under $H(x,0,t)$ at each $t$.  In fact, using Eqs.\,\eqref{Fpiconnection2}, \eqref{defqxb}, \eqref{bperp2}, one finds
\begin{subequations}
\begin{align}
\langle |\vec{b}_\perp|^2 \rangle & = \int_{-1}^1 \rule{-1ex}{0ex}dx \,
\langle |\vec{b}_\perp(x;\zeta)|^2 \rangle =
\int_0^\infty \rule{-1ex}{0ex}  d^2 |\vec{b}_\perp|\,|\vec{b}_\perp|^2\,{\mathpzc d}_\pi(|\vec{b}_\perp|) \,,\rule{-2ex}{0ex} \\
{\mathpzc d}_\pi(|\vec{b}_\perp|) &= \int_0^\infty d{\Delta} \,{\Delta}\, J_0(|\vec{b}_\perp| {\Delta}) F_\pi({\Delta}^2)\,,
\end{align}
\end{subequations}
and hence, with $F_\pi(t)$ in Eq.\,\eqref{eqFpit}, $\langle |\vec{b}_\perp|^2 \rangle = (0.52\,{\rm fm})^2$.  For the reasons just explained, this is also the value obtained with Eq.\,\eqref{Hvtau0}.  [Note that $F_\pi({\mathpzc t}) = 1/(1 + {\mathpzc t} r_\pi^2/6)$, where $r_\pi$ is the pion's electric charge radius, yields $\langle |\vec{b}_\perp|^2 \rangle = (2/3) r_\pi^2 = (0.55\,{\rm fm})^2$ (empirically \cite{Beringer:1900zz} $r_\pi = 0.672 \pm 0.008\,$fm).]  It is natural that the mean-squared transverse extent of the dressed-quarks within a pion should be commensurate with the length-scale associated with light-quark confinement realised through a violation of reflection positivity [see, e.g., Sect.\,2.2 in Ref.\,\cite{Cloet:2013jya}].


\smallskip

\noindent\textbf{6.$\;$Connection between the pion's GPD and its PDA}.
We have hitherto focused on $\xi=0$, with $t$ ranging over all spacelike momentum transfer but we will now consider another extreme; viz., $\xi=1$, with $t\to 0$.  In this kinematic scenario the dressed-quark GPD is obtained from Eq.\,\eqref{eq:TriangleDiagrams} by inserting $P=-\Delta/2$, so that the incoming pion momentum is $p_\pi=-\Delta$ and the outgoing $p_\pi^\prime \to 0$, to obtain [$u=(1+x)/2$]:
\begin{align}
\nonumber
& \rule{-1.3ex}{0ex} 2 H_{\pi}^{\rm v}(x,1,0)
 = N_c \mathrm{tr} \int_{d\ell}\delta_n^{up_\pi}(\ell)  \, S(\ell_{p_\pi})  \\
&\rule{-1.3ex}{0ex}\times
i{\Gamma}_\pi(\ell_{p_\pi};p_\pi^\prime )\,
S(\ell_{p_\pi}) \,in\cdot\Gamma(\ell_{p_\pi},\ell) \, S(\ell)
 i\Gamma_\pi(\ell;p_\pi)\,,
\label{eq:TriangleDiagramxi1}
\end{align}
where we have chosen $\eta=1$ for simplicity and shifted the integration variable $\ell\to\ell+\Delta/2$.  The final result is obtained by taking the limit $p_\pi^\prime\to 0$ and $\Delta\to 0$, in that order.

In proceeding, let us maintain, as described in connection with Eq.\,\eqref{eq:TriangleDiagrams}, that every element in Eq.\,\eqref{eq:TriangleDiagramxi1} is computed in the symmetry-preserving RL truncation, in which case the following two DCSB-induced soft-pion theorems are \begin{equation}
\label{softpion}
\begin{array}{rcl}
2 f_\pi \Gamma_\pi(\ell_{p_\pi}; p_\pi^\prime ) &\stackrel{p_\pi^\prime \simeq 0}{\approx}&
p^\prime_{\pi\mu} \Gamma_{5\mu}(\ell_{p_\pi},\ell_{p_\pi})\,, \\
2r_\pi\Gamma_\pi(\ell;p_\pi) &\stackrel{p_\pi \simeq 0}{\approx}&
\Gamma_{5}(\ell,\ell_{p_\pi})\,,
\end{array}
\end{equation}
where $f_\pi$ and $r_\pi$ are, respectively, the residues of the pion pole in the inhomogeneous pseudovector and pseudoscalar vertices.  Now, using Eqs.\,\eqref{softpion} in concert with a straightforward generalisation of the rainbow-ladder axial-vector Ward-Takahashi identity displayed in Fig.\,1 of Ref.\,\cite{Holl:2005vu}, Eq.\,\eqref{eq:TriangleDiagramxi1} simplifies:
{\allowdisplaybreaks
\begin{align}
\nonumber
&  2 H_{\pi}^{\rm v}(x,1,0)
 \stackrel{\Delta^2\simeq 0}{=} \frac{\Delta^2}{4f_\pi r_\pi}
 N_c\mathrm{tr} \rule{-1ex}{0ex}\int_{d\ell}\,\delta_n^{up_\pi}(\ell)\, S(\ell_{p_\pi}) \\
\nonumber
&\quad \times
p^\prime_{\pi\mu} i\Gamma_{5\mu}(\ell_{p_\pi},\ell_{p_\pi})
S(\ell_{p_\pi}) \,in\cdot\Gamma(\ell_{p_\pi},\ell) \, S(\ell)
 i\Gamma_{5}(\ell,\ell_{p_\pi})\\
\nonumber
&\stackrel{\Delta^2\simeq 0}{=} \frac{\Delta^2}{4f_\pi r_\pi} N_c \mathrm{tr} \rule{-1ex}{0ex}\int_{d\ell}\, \delta_n^{up_\pi}(\ell)
\left[ Z_2\gamma_5 \gamma\cdot n S(\ell)\Gamma_5(\ell,\ell_{p_\pi}) S(\ell_{p_\pi}) \right.
 \\
&
 \quad \quad \left.
 +Z_4 \mathbf{1} S(\ell_{p_\pi}) n\cdot \Gamma(\ell_{p_\pi},\ell) S(\ell)
\right]\,, \label{Hphi1}
\end{align}}
\hspace*{-0.4\parindent}where $Z_{2,4}$ are, respectively, renormalisation constants for the inhomogeneous vector and pseudoscalar vertices.  The last term on the rhs of Eq.\,\eqref{Hphi1} is zero because \cite{Bhagwat:2006py} the inhomogeneous vector vertex does not contain a zero mass pole in the presence of DCSB and, moreover, in the chiral limit a vector probe cannot couple to a $J^{PC}=0^{++}$ final state.  Consequently, one has
\begin{align}
\nonumber
2 H_{\pi}^{\rm v}(x,1,0)& =
\frac{1}{2f_\pi}N_c\mathrm{tr}Z_2 \rule{-1ex}{0ex}\int_{d\ell} \delta_n^{up_\pi}(\ell)\,\gamma_5 \gamma\cdot n S(\ell) \Gamma_\pi(\ell; p_\pi ) S(\ell_{p_\pi})\\
& = \frac{1}{2} \varphi_\pi(u)\,,
\end{align}
where $\varphi_\pi(u)$ is the pion's valence dressed-quark parton distribution amplitude (PDA).  Using a change of integration variable and the charge conjugation properties of the elements in Eq.\,\eqref{eq:TriangleDiagramxi1}, it is straightforward to show $H_{\pi}^{\rm v}(x,1,0)=H^{\rm v}_{\pi}(-x,1,0)$.  Hence, the analysis in this Section is the derivation in rainbow-ladder truncation
of a general result \cite{Polyakov:1998ze,Polyakov:1999gs}:
\begin{equation}
\label{GPDPDA}
\begin{array}{c}
H^{I=1}_{\pi}(2 u -1 ,1,0) = \frac{1}{2} \varphi_\pi(u)\,,\; u\in[0,1]\,.
\end{array}
\end{equation}

Employing a similar procedure, one can show $H_\pi^{\rm v}(x,-1,t) =H_\pi^{\rm v}(x,1,t)$, which is a particular case of the general property $H^{q}(x,\xi,t) =H^{q}(x,-\xi,t)$ that follows from time reversal invariance.  Thus, in a fully consistent rainbow-ladder truncation, any valid correction to Eq.\,\eqref{eq:TriangleDiagrams} must vanish at $\xi=\pm 1$.

\smallskip

\noindent\textbf{7.$\;$Conclusion and prospects}.
We described a calculation of the pion's valence dressed-quark generalised parton distribution (GPD), $H_\pi^{\rm v}(x,\xi,t)$, within the context of a rainbow-ladder (RL) truncation of QCD's Dyson-Schwinger equations.  This framework is useful at an hadronic scale because it provides a description of hadrons via a dressed-quark basis, the accuracy of which in any given channel is knowable \emph{a priori}.  Our analysis focused primarily on the case of zero skewness ($\xi=0$) but we also capitalised on the symmetry-preserving character of the RL truncation in order to demonstrate a known relationship between $H_\pi^{\rm v}(x,\xi=\pm 1,t)$ and the pion's valence-quark parton distribution amplitude [Eq.\,\eqref{GPDPDA}].

Drawing analogy with the pion's valence dressed-quark distribution function, we argued that the impulse-approximation used hitherto to define the pion's valence GPD is generally invalid owing to omission of contributions from the gluons which bind dressed-quarks into the pion.  We used a simple correction [Eq.\eqref{HCorrection}], valid in the neighbourhood of $\xi=0$, $t=0$, in order to identify a practicable improvement to the approximation for $H_\pi^{\rm v}(x,0,t)$.  Expressing the result as the Radon transform of a single amplitude, we were able to isolate and remove those terms which produce unphysical behaviour, such as violations of positivity by the $\xi=0$ GPD.  The resulting, ameliorated Radon amplitude yields a form for $H_\pi^{\rm v}(x,0,t)$ [Eqs.\,\eqref{Hgeneral}, \eqref{Cfinal}] which is consistent with significantly more known constraints than is the result produced by the impulse approximation alone.

The results obtained in this way for $H_\pi^{\rm v}(x,0,t)$, $q_\pi^{\rm v}(x,|\vec{b}_\perp|)$ [Figs.\,\ref{figHx0tz0z2}, \ref{figqbx2}] provide a qualitatively sound picture of the dressed-quark structure of the pion at an hadronic scale.  Using leading-order expressions, we evolved these distributions to a scale $\zeta=2\,$GeV.  All features of the resulting valence quark GPDs may be intuitively understood and hence the distributions should serve as an elementary but reasonable guide in the planning and interpretation of relevant experiments at existing or anticipated \cite{Accardi:2012qut} facilities, which could plausibly involve deeply-virtual Compton scattering on pions in a nucleon's meson cloud.

Notwithstanding the simplicity of the framework employed herein, a merit of the approach is its potential to \emph{compute} features of hadron GPDs on the valence-quark domain and relate them directly to properties of QCD.  This capacity has already been demonstrated in the simpler case of the pion's valence parton distribution function \cite{Chang:2014lva}.  One may begin to realise that potential by using more realistic forms for the dressed-propagators and -vertices that appear in the RL truncation analysis and, perhaps more importantly, uncovering the amendment to impulse approximation which is required in order to extend the validity of the RL truncation to the entire kinematic domain of $\xi$ and $t$.



\smallskip

\noindent\textbf{Acknowledgments}.
We thank
A.~Besse,
I.\,C.~Clo\"et,
D.\,M\"uller,
P.~From\-holz,
C.~Keppel,
P.~Kroll,
J.-Ph.~Lansberg
C.~Lorc\'e,
J.~Segovia
and
S.~Wallon
for valuable discussions.
CM, LC, HM, CDR and JR-Q are grateful for the chance to participate in the workshop ``Many Manifestations of Nonperturbative QCD under the Southern Cross'', Ubatuba, S\~ao Paulo, where significant parts of this work were first presented and improvements discussed.
CDR acknowledges support from an \emph{International Fellow Award} from the Helmholtz Association; and
this research was otherwise supported by:
Commissariat \`a l'Energie Atomique;
%
JRA ``Study of Strongly Interacting Matter'' (Grant Agreement no.\,283286, HadronPhysics3)
%
under the 7th E.U. F.P.;
GDR 3034 PH-QCD; 
ANR-12-MONU-0008-01 ``PARTONS'';
University of Adelaide and Australian Research Council through grant no.~FL0992247;
Spanish ministry Research Project FPA2011-23781;
U.S.\ Department of Energy, Office of Science, Office of Nuclear Physics, contract no.~DE-AC02-06CH11357;
and For\-schungs\-zentrum J\"ulich GmbH.

\vspace*{-2ex}



\begin{thebibliography}{69}
\expandafter\ifx\csname natexlab\endcsname\relax\def\natexlab#1{#1}\fi
\providecommand{\bibinfo}[2]{#2}
\ifx\xfnm\relax \def\xfnm[#1]{\unskip,\space#1}\fi
\bibitem[{Friedman(1991)}]{Friedman:1991nq}
\bibinfo{author}{J.~I. Friedman}, \bibinfo{journal}{Rev. Mod. Phys.}
  \bibinfo{volume}{63} (\bibinfo{year}{1991}) \bibinfo{pages}{615--629}.
\bibitem[{Taylor(1991)}]{Taylor:1991ew}
\bibinfo{author}{R.~E. Taylor}, \bibinfo{journal}{Rev. Mod. Phys.}
  \bibinfo{volume}{63} (\bibinfo{year}{1991}) \bibinfo{pages}{573--595}.
\bibitem[{Kendall(1991)}]{Kendall:1991np}
\bibinfo{author}{H.~W. Kendall}, \bibinfo{journal}{Rev. Mod. Phys.}
  \bibinfo{volume}{63} (\bibinfo{year}{1991}) \bibinfo{pages}{597--614}.
\bibitem[{Keister and Polyzou(1991)}]{Keister:1991sb}
\bibinfo{author}{B.~D. Keister}, \bibinfo{author}{W.~N. Polyzou},
  \bibinfo{journal}{Adv. Nucl. Phys.} \bibinfo{volume}{20}
  (\bibinfo{year}{1991}) \bibinfo{pages}{225--479}.
\bibitem[{Coester(1992)}]{Coester:1992cg}
\bibinfo{author}{F.~Coester}, \bibinfo{journal}{Prog. Part. Nucl. Phys.}
  \bibinfo{volume}{29} (\bibinfo{year}{1992}) \bibinfo{pages}{1--32}.
\bibitem[{Brodsky et~al.(1998)Brodsky, Pauli, and Pinsky}]{Brodsky:1997de}
\bibinfo{author}{S.~J. Brodsky}, \bibinfo{author}{H.-C. Pauli},
  \bibinfo{author}{S.~S. Pinsky}, \bibinfo{journal}{Phys. Rept.}
  \bibinfo{volume}{301} (\bibinfo{year}{1998}) \bibinfo{pages}{299--486}.
\bibitem[{Myhrer and Thomas(2010)}]{Myhrer:2009uq}
\bibinfo{author}{F.~Myhrer}, \bibinfo{author}{A.~W. Thomas},
  \bibinfo{journal}{J. Phys. G} \bibinfo{volume}{37} (\bibinfo{year}{2010})
  \bibinfo{pages}{023101}.
\bibitem[{Jimenez-Delgado et~al.(2013)Jimenez-Delgado, Melnitchouk, and
  Owens}]{Jimenez-Delgado:2013sma}
\bibinfo{author}{P.~Jimenez-Delgado}, \bibinfo{author}{W.~Melnitchouk},
  \bibinfo{author}{J.~Owens}, \bibinfo{journal}{J. Phys. G}
  \bibinfo{volume}{40} (\bibinfo{year}{2013}) \bibinfo{pages}{093102}.
\bibitem[{Dittes et~al.(1988)Dittes, M{\"u}ller, Robaschik, Geyer, and
  Ho{\v{r}}ej{\v{s}}i}]{Dittes:1988xz}
\bibinfo{author}{F.~M. Dittes}, \bibinfo{author}{D.~M{\"u}ller},
  \bibinfo{author}{D.~Robaschik}, \bibinfo{author}{B.~Geyer},
  \bibinfo{author}{J.~Ho{\v{r}}ej{\v{s}}i}, \bibinfo{journal}{Phys. Lett. B}
  \bibinfo{volume}{209} (\bibinfo{year}{1988}) \bibinfo{pages}{325--329}.
\bibitem[{Mueller et~al.(1994)Mueller, Robaschik, Geyer, Dittes, and
  Ho\v{r}ej\v{s}i}]{Mueller:1998fv}
\bibinfo{author}{D.~Mueller}, \bibinfo{author}{D.~Robaschik},
  \bibinfo{author}{B.~Geyer}, \bibinfo{author}{F.~M. Dittes},
  \bibinfo{author}{J.~Ho\v{r}ej\v{s}i}, \bibinfo{journal}{Fortschr. Phys.}
  \bibinfo{volume}{42} (\bibinfo{year}{1994}) \bibinfo{pages}{101}.
\bibitem[{Radyushkin(1996)}]{Radyushkin:1996nd}
\bibinfo{author}{A.~Radyushkin}, \bibinfo{journal}{Phys. Lett. B}
  \bibinfo{volume}{380} (\bibinfo{year}{1996}) \bibinfo{pages}{417--425}.
\bibitem[{Ji(1997)}]{Ji:1996nm}
\bibinfo{author}{X.-D. Ji}, \bibinfo{journal}{Phys. Rev. D}
  \bibinfo{volume}{55} (\bibinfo{year}{1997}) \bibinfo{pages}{7114--7125}.
\bibitem[{Burkardt(2003)}]{Burkardt:2002hr}
\bibinfo{author}{M.~Burkardt}, \bibinfo{journal}{Int. J. Mod. Phys. A}
  \bibinfo{volume}{18} (\bibinfo{year}{2003}) \bibinfo{pages}{173--208}.
\bibitem[{Diehl(2003)}]{Diehl:2003ny}
\bibinfo{author}{M.~Diehl}, \bibinfo{journal}{Phys. Rept.}
  \bibinfo{volume}{388} (\bibinfo{year}{2003}) \bibinfo{pages}{41--277}.
\bibitem[{Belitsky and Radyushkin(2005)}]{Belitsky:2005qn}
\bibinfo{author}{A.~Belitsky}, \bibinfo{author}{A.~Radyushkin},
  \bibinfo{journal}{Phys. Rept.} \bibinfo{volume}{418} (\bibinfo{year}{2005})
  \bibinfo{pages}{1--387}.
\bibitem[{Boffi and Pasquini(2007)}]{Boffi:2007yc}
\bibinfo{author}{S.~Boffi}, \bibinfo{author}{B.~Pasquini},
  \bibinfo{journal}{Riv. Nuovo Cim.} \bibinfo{volume}{30}
  (\bibinfo{year}{2007}) \bibinfo{pages}{387--448}.
\bibitem[{Guidal et~al.(2013)Guidal, Moutarde, and
  Vanderhaeghen}]{Guidal:2013rya}
\bibinfo{author}{M.~Guidal}, \bibinfo{author}{H.~Moutarde},
  \bibinfo{author}{M.~Vanderhaeghen}, \bibinfo{journal}{Rept. Prog. Phys.}
  \bibinfo{volume}{76} (\bibinfo{year}{2013}) \bibinfo{pages}{066202}.
\bibitem[{Radyushkin(1999{\natexlab{a}})}]{Radyushkin:1998es}
\bibinfo{author}{A.~Radyushkin}, \bibinfo{journal}{Phys. Rev. D}
  \bibinfo{volume}{59} (\bibinfo{year}{1999}{\natexlab{a}})
  \bibinfo{pages}{014030}.
\bibitem[{Radyushkin(1999{\natexlab{b}})}]{Radyushkin:1998bz}
\bibinfo{author}{A.~Radyushkin}, \bibinfo{journal}{Phys. Lett. B}
  \bibinfo{volume}{449} (\bibinfo{year}{1999}{\natexlab{b}})
  \bibinfo{pages}{81--88}.
\bibitem[{Teryaev(2001)}]{Teryaev:2001qm}
\bibinfo{author}{O.~Teryaev}, \bibinfo{journal}{Phys. Lett. B}
  \bibinfo{volume}{510} (\bibinfo{year}{2001}) \bibinfo{pages}{125--132}.
\bibitem[{Musatov and Radyushkin(2000)}]{Musatov:1999xp}
\bibinfo{author}{I.~Musatov}, \bibinfo{author}{A.~Radyushkin},
  \bibinfo{journal}{Phys. Rev. D} \bibinfo{volume}{61} (\bibinfo{year}{2000})
  \bibinfo{pages}{074027}.
\bibitem[{Mezrag et~al.(2013)Mezrag, Moutarde, and
  Sabati{\'e}}]{Mezrag:2013mya}
\bibinfo{author}{C.~Mezrag}, \bibinfo{author}{H.~Moutarde},
  \bibinfo{author}{F.~Sabati{\'e}}, \bibinfo{journal}{Phys. Rev. D}
  \bibinfo{volume}{88} (\bibinfo{year}{2013}) \bibinfo{pages}{014001}.
\bibitem[{Kumericki et~al.(2008)Kumericki, Mueller, and
  Passek-Kumericki}]{Kumericki:2008di}
\bibinfo{author}{K.~Kumericki}, \bibinfo{author}{D.~Mueller},
  \bibinfo{author}{K.~Passek-Kumericki}, \bibinfo{journal}{Eur. Phys. J. C}
  \bibinfo{volume}{58} (\bibinfo{year}{2008}) \bibinfo{pages}{193--215}.
\bibitem[{Chang et~al.(2014)Chang, Mezrag, Moutarde, Roberts,
  Rodr{\'{\i}}guez-Quintero, and Tandy}]{Chang:2014lva}
\bibinfo{author}{L.~Chang}, \bibinfo{author}{C.~Mezrag},
  \bibinfo{author}{H.~Moutarde}, \bibinfo{author}{C.~D. Roberts},
  \bibinfo{author}{J.~Rodr{\'{\i}}guez-Quintero}, \bibinfo{author}{P.~C.
  Tandy}, \bibinfo{journal}{Phys. Lett. B} \bibinfo{volume}{737}
  (\bibinfo{year}{2014}) \bibinfo{pages}{23–29}.
\bibitem[{Maris et~al.(1998)Maris, Roberts, and Tandy}]{Maris:1997hd}
\bibinfo{author}{P.~Maris}, \bibinfo{author}{C.~D. Roberts},
  \bibinfo{author}{P.~C. Tandy}, \bibinfo{journal}{Phys. Lett. B}
  \bibinfo{volume}{420} (\bibinfo{year}{1998}) \bibinfo{pages}{267--273}.
\bibitem[{Qin et~al.(2014)Qin, Roberts, and Schmidt}]{Qin:2014vya}
\bibinfo{author}{S.-X. Qin}, \bibinfo{author}{C.~D. Roberts},
  \bibinfo{author}{S.~M. Schmidt}, \bibinfo{journal}{Phys. Lett. B}
  \bibinfo{volume}{733} (\bibinfo{year}{2014}) \bibinfo{pages}{202--208}.
\bibitem[{Chang et~al.(2011)Chang, Roberts, and Tandy}]{Chang:2011vu}
\bibinfo{author}{L.~Chang}, \bibinfo{author}{C.~D. Roberts},
  \bibinfo{author}{P.~C. Tandy}, \bibinfo{journal}{Chin. J. Phys.}
  \bibinfo{volume}{49} (\bibinfo{year}{2011}) \bibinfo{pages}{955--1004}.
\bibitem[{Bashir et~al.(2012)}]{Bashir:2012fs}
\bibinfo{author}{A.~Bashir}, et~al., \bibinfo{journal}{Commun. Theor. Phys.}
  \bibinfo{volume}{58} (\bibinfo{year}{2012}) \bibinfo{pages}{79--134}.
\bibitem[{Clo{\"e}t and Roberts(2014)}]{Cloet:2013jya}
\bibinfo{author}{I.~C. Clo{\"e}t}, \bibinfo{author}{C.~D. Roberts},
  \bibinfo{journal}{Prog. Part. Nucl. Phys.} \bibinfo{volume}{77}
  (\bibinfo{year}{2014}) \bibinfo{pages}{1--69}.
\bibitem[{Efremov and Radyushkin(1980)}]{Efremov:1979qk}
\bibinfo{author}{A.~V. Efremov}, \bibinfo{author}{A.~V. Radyushkin},
  \bibinfo{journal}{Phys. Lett. B} \bibinfo{volume}{94} (\bibinfo{year}{1980})
  \bibinfo{pages}{245--250}.
\bibitem[{Lepage and Brodsky(1980)}]{Lepage:1980fj}
\bibinfo{author}{G.~P. Lepage}, \bibinfo{author}{S.~J. Brodsky},
  \bibinfo{journal}{Phys. Rev. D} \bibinfo{volume}{22} (\bibinfo{year}{1980})
  \bibinfo{pages}{2157--2198}.
\bibitem[{Dokshitzer(1977)}]{Dokshitzer:1977}
\bibinfo{author}{Y.~L. Dokshitzer}, \bibinfo{journal}{Sov. Phys. JETP}
  \bibinfo{volume}{46} (\bibinfo{year}{1977}) \bibinfo{pages}{641--653}.
\bibitem[{Gribov and Lipatov(1972)}]{Gribov:1972}
\bibinfo{author}{V.~N. Gribov}, \bibinfo{author}{L.~N. Lipatov},
  \bibinfo{journal}{Sov. J. Nucl. Phys.} \bibinfo{volume}{15}
  (\bibinfo{year}{1972}) \bibinfo{pages}{438--450}.
\bibitem[{Lipatov(1975)}]{Lipatov:1974qm}
\bibinfo{author}{L.~N. Lipatov}, \bibinfo{journal}{Sov. J. Nucl. Phys.}
  \bibinfo{volume}{20} (\bibinfo{year}{1975}) \bibinfo{pages}{94--102}.
\bibitem[{Altarelli and Parisi(1977)}]{Altarelli:1977}
\bibinfo{author}{G.~Altarelli}, \bibinfo{author}{G.~Parisi},
  \bibinfo{journal}{Nucl. Phys. B} \bibinfo{volume}{126} (\bibinfo{year}{1977})
  \bibinfo{pages}{298}.
\bibitem[{Bashir et~al.(2008)Bashir, Raya, Clo{\"e}t, and
  Roberts}]{Bashir:2008fk}
\bibinfo{author}{A.~Bashir}, \bibinfo{author}{A.~Raya}, \bibinfo{author}{I.~C.
  Clo{\"e}t}, \bibinfo{author}{C.~D. Roberts}, \bibinfo{journal}{Phys.\ Rev.\
  C} \bibinfo{volume}{78} (\bibinfo{year}{2008}) \bibinfo{pages}{055201}.
\bibitem[{Bashir et~al.(2009)Bashir, Raya, S{\'a}nchez-Madrigal, and
  Roberts}]{Bashir:2009fv}
\bibinfo{author}{A.~Bashir}, \bibinfo{author}{A.~Raya},
  \bibinfo{author}{S.~S{\'a}nchez-Madrigal}, \bibinfo{author}{C.~D. Roberts},
  \bibinfo{journal}{Few Body Syst.} \bibinfo{volume}{46} (\bibinfo{year}{2009})
  \bibinfo{pages}{229--237}.
\bibitem[{Kopeliovich et~al.(2013)Kopeliovich, Schmidt, and
  Siddikov}]{Kopeliovich:2011rv}
\bibinfo{author}{B.~Kopeliovich}, \bibinfo{author}{I.~Schmidt},
  \bibinfo{author}{M.~Siddikov}, \bibinfo{journal}{Nucl. Phys. A}
  \bibinfo{volume}{918} (\bibinfo{year}{2013}) \bibinfo{pages}{41--60}.
\bibitem[{Mezrag et~al.(p ph)Mezrag, Moutarde, Rodr{\'i}guez-Quintero, and
  Sabati{\'e}}]{Mezrag:2014tva}
\bibinfo{author}{C.~Mezrag}, \bibinfo{author}{H.~Moutarde},
  \bibinfo{author}{J.~Rodr{\'i}guez-Quintero}, \bibinfo{author}{F.~Sabati{\'e}}
   (\bibinfo{year}{arXiv:1406.7425 [hep-ph]}). \bibinfo{note}{{\emph{Towards a
  Pion Generalized Parton Distribution Model from Dyson-Schwinger Equations}}}.
\bibitem[{Burkardt(2000)}]{Burkardt:2000za}
\bibinfo{author}{M.~Burkardt}, \bibinfo{journal}{Phys. Rev. D}
  \bibinfo{volume}{62} (\bibinfo{year}{2000}) \bibinfo{pages}{071503}.
\bibitem[{Diehl et~al.(2001)Diehl, Feldmann, Jakob, and Kroll}]{Diehl:2000xz}
\bibinfo{author}{M.~Diehl}, \bibinfo{author}{T.~Feldmann},
  \bibinfo{author}{R.~Jakob}, \bibinfo{author}{P.~Kroll},
  \bibinfo{journal}{Nucl. Phys. B} \bibinfo{volume}{596} (\bibinfo{year}{2001})
  \bibinfo{pages}{33--65}.
\bibitem[{Chang et~al.(2013)Chang, Clo{\"e}t, Roberts, Schmidt, and
  Tandy}]{Chang:2013nia}
\bibinfo{author}{L.~Chang}, \bibinfo{author}{I.~C. Clo{\"e}t},
  \bibinfo{author}{C.~D. Roberts}, \bibinfo{author}{S.~M. Schmidt},
  \bibinfo{author}{P.~C. Tandy}, \bibinfo{journal}{Phys. Rev. Lett.}
  \bibinfo{volume}{111} (\bibinfo{year}{2013}) \bibinfo{pages}{141802}.
\bibitem[{Munczek(1995)}]{Munczek:1994zz}
\bibinfo{author}{H.~J. Munczek}, \bibinfo{journal}{Phys. Rev. D}
  \bibinfo{volume}{52} (\bibinfo{year}{1995}) \bibinfo{pages}{4736--4740}.
\bibitem[{Bender et~al.(1996)Bender, Roberts, and von Smekal}]{Bender:1996bb}
\bibinfo{author}{A.~Bender}, \bibinfo{author}{C.~D. Roberts},
  \bibinfo{author}{L.~von Smekal}, \bibinfo{journal}{Phys. Lett. B}
  \bibinfo{volume}{380} (\bibinfo{year}{1996}) \bibinfo{pages}{7--12}.
\bibitem[{Hecht et~al.(2001)Hecht, Roberts, and Schmidt}]{Hecht:2000xa}
\bibinfo{author}{M.~B. Hecht}, \bibinfo{author}{C.~D. Roberts},
  \bibinfo{author}{S.~M. Schmidt}, \bibinfo{journal}{Phys. Rev. C}
  \bibinfo{volume}{63} (\bibinfo{year}{2001}) \bibinfo{pages}{025213}.
\bibitem[{Nguyen et~al.(2011)Nguyen, Bashir, Roberts, and
  Tandy}]{Nguyen:2011jy}
\bibinfo{author}{T.~Nguyen}, \bibinfo{author}{A.~Bashir},
  \bibinfo{author}{C.~D. Roberts}, \bibinfo{author}{P.~C. Tandy},
  \bibinfo{journal}{Phys. Rev. C} \bibinfo{volume}{83} (\bibinfo{year}{2011})
  \bibinfo{pages}{062201(R)}.
\bibitem[{Tiburzi and Miller(2003)}]{Tiburzi:2002tq}
\bibinfo{author}{B.~Tiburzi}, \bibinfo{author}{G.~Miller},
  \bibinfo{journal}{Phys. Rev. D} \bibinfo{volume}{67} (\bibinfo{year}{2003})
  \bibinfo{pages}{113004}.
\bibitem[{Broniowski and Ruiz~Arriola(2003)}]{Broniowski:2003rp}
\bibinfo{author}{W.~Broniowski}, \bibinfo{author}{E.~Ruiz~Arriola},
  \bibinfo{journal}{Phys. Lett. B} \bibinfo{volume}{574} (\bibinfo{year}{2003})
  \bibinfo{pages}{57--64}.
\bibitem[{Ji et~al.(2006)Ji, Mishchenko, and Radyushkin}]{Ji:2006ea}
\bibinfo{author}{C.-R. Ji}, \bibinfo{author}{Y.~Mishchenko},
  \bibinfo{author}{A.~Radyushkin}, \bibinfo{journal}{Phys. Rev. D}
  \bibinfo{volume}{73} (\bibinfo{year}{2006}) \bibinfo{pages}{114013}.
\bibitem[{Broniowski et~al.(2008)Broniowski, Ruiz~Arriola, and
  Golec-Biernat}]{Broniowski:2007si}
\bibinfo{author}{W.~Broniowski}, \bibinfo{author}{E.~Ruiz~Arriola},
  \bibinfo{author}{K.~Golec-Biernat}, \bibinfo{journal}{Phys. Rev. D}
  \bibinfo{volume}{77} (\bibinfo{year}{2008}) \bibinfo{pages}{034023}.
\bibitem[{Frederico et~al.(2009)Frederico, Pace, Pasquini, and
  Salme}]{Frederico:2009fk}
\bibinfo{author}{T.~Frederico}, \bibinfo{author}{E.~Pace},
  \bibinfo{author}{B.~Pasquini}, \bibinfo{author}{G.~Salme},
  \bibinfo{journal}{Phys.Rev.} \bibinfo{volume}{D80} (\bibinfo{year}{2009})
  \bibinfo{pages}{054021}.
\bibitem[{Chang et~al.(2013)Chang, Clo{\"e}t, Cobos-Martinez, Roberts, Schmidt,
  and Tandy}]{Chang:2013pq}
\bibinfo{author}{L.~Chang}, \bibinfo{author}{I.~C. Clo{\"e}t},
  \bibinfo{author}{J.~J. Cobos-Martinez}, \bibinfo{author}{C.~D. Roberts},
  \bibinfo{author}{S.~M. Schmidt}, \bibinfo{author}{P.~C. Tandy},
  \bibinfo{journal}{Phys. Rev. Lett.} \bibinfo{volume}{110}
  (\bibinfo{year}{2013}) \bibinfo{pages}{132001}.
\bibitem[{Roberts et~al.(2011)Roberts, Bashir, Guti{\'e}rrez-Guerrero, Roberts,
  and Wilson}]{Roberts:2011wy}
\bibinfo{author}{H.~L.~L. Roberts}, \bibinfo{author}{A.~Bashir},
  \bibinfo{author}{L.~X. Guti{\'e}rrez-Guerrero}, \bibinfo{author}{C.~D.
  Roberts}, \bibinfo{author}{D.~J. Wilson}, \bibinfo{journal}{Phys. Rev. C}
  \bibinfo{volume}{83} (\bibinfo{year}{2011}) \bibinfo{pages}{065206}.
\bibitem[{Huber et~al.(2008)}]{Huber:2008id}
\bibinfo{author}{G.~Huber}, et~al., \bibinfo{journal}{Phys. Rev. C}
  \bibinfo{volume}{78} (\bibinfo{year}{2008}) \bibinfo{pages}{045203}.
\bibitem[{Farrar and Jackson(1979)}]{Farrar:1979aw}
\bibinfo{author}{G.~R. Farrar}, \bibinfo{author}{D.~R. Jackson},
  \bibinfo{journal}{Phys. Rev. Lett.} \bibinfo{volume}{43}
  (\bibinfo{year}{1979}) \bibinfo{pages}{246--249}.
\bibitem[{Maris and Roberts(1998)}]{Maris:1998hc}
\bibinfo{author}{P.~Maris}, \bibinfo{author}{C.~D. Roberts},
  \bibinfo{journal}{Phys. Rev. C} \bibinfo{volume}{58} (\bibinfo{year}{1998})
  \bibinfo{pages}{3659--3665}.
\bibitem[{Bhagwat et~al.(2003)Bhagwat, Pichowsky, Roberts, and
  Tandy}]{Bhagwat:2003vw}
\bibinfo{author}{M.~Bhagwat}, \bibinfo{author}{M.~Pichowsky},
  \bibinfo{author}{C.~Roberts}, \bibinfo{author}{P.~Tandy},
  \bibinfo{journal}{Phys. Rev. C} \bibinfo{volume}{68} (\bibinfo{year}{2003})
  \bibinfo{pages}{015203}.
\bibitem[{Bowman et~al.(2005)}]{Bowman:2005vx}
\bibinfo{author}{P.~O. Bowman}, et~al., \bibinfo{journal}{Phys. Rev. D}
  \bibinfo{volume}{71} (\bibinfo{year}{2005}) \bibinfo{pages}{054507}.
\bibitem[{Bhagwat and Tandy(2006)}]{Bhagwat:2006tu}
\bibinfo{author}{M.~S. Bhagwat}, \bibinfo{author}{P.~C. Tandy},
  \bibinfo{journal}{AIP Conf. Proc.} \bibinfo{volume}{842}
  (\bibinfo{year}{2006}) \bibinfo{pages}{225--227}.
\bibitem[{Gluck et~al.(1999)Gluck, Reya, and Schienbein}]{Gluck:1999xe}
\bibinfo{author}{M.~Gluck}, \bibinfo{author}{E.~Reya},
  \bibinfo{author}{I.~Schienbein}, \bibinfo{journal}{Eur. Phys. J. C}
  \bibinfo{volume}{10} (\bibinfo{year}{1999}) \bibinfo{pages}{313--317}.
\bibitem[{Georgi and Politzer(1974)}]{Georgi:1951sr}
\bibinfo{author}{H.~Georgi}, \bibinfo{author}{H.~D. Politzer},
  \bibinfo{journal}{Phys. Rev. D} \bibinfo{volume}{9} (\bibinfo{year}{1974})
  \bibinfo{pages}{416--420}.
\bibitem[{Gross and Wilczek(1974)}]{Gross:1974cs}
\bibinfo{author}{D.~Gross}, \bibinfo{author}{F.~Wilczek},
  \bibinfo{journal}{Phys. Rev. D} \bibinfo{volume}{9} (\bibinfo{year}{1974})
  \bibinfo{pages}{980--993}.
\bibitem[{Politzer(1974)}]{Politzer:1974fr}
\bibinfo{author}{H.~D. Politzer}, \bibinfo{journal}{Phys. Rept.}
  \bibinfo{volume}{14} (\bibinfo{year}{1974}) \bibinfo{pages}{129--180}.
\bibitem[{Beringer et~al.(2012)}]{Beringer:1900zz}
\bibinfo{author}{J.~Beringer}, et~al., \bibinfo{journal}{Phys. Rev. D}
  \bibinfo{volume}{86} (\bibinfo{year}{2012}) \bibinfo{pages}{010001}.
\bibitem[{H{\"o}ll et~al.(2005)H{\"o}ll, Krassnigg, Maris, Roberts, and
  Wright}]{Holl:2005vu}
\bibinfo{author}{A.~H{\"o}ll}, \bibinfo{author}{A.~Krassnigg},
  \bibinfo{author}{P.~Maris}, \bibinfo{author}{C.~D. Roberts},
  \bibinfo{author}{S.~V. Wright}, \bibinfo{journal}{Phys. Rev. C}
  \bibinfo{volume}{71} (\bibinfo{year}{2005}) \bibinfo{pages}{065204}.
\bibitem[{Bhagwat et~al.(2007)Bhagwat, H{\"o}ll, Krassnigg, and
  Roberts}]{Bhagwat:2006py}
\bibinfo{author}{M.~S. Bhagwat}, \bibinfo{author}{A.~H{\"o}ll},
  \bibinfo{author}{A.~Krassnigg}, \bibinfo{author}{C.~D. Roberts},
  \bibinfo{journal}{Nucl. Phys. A} \bibinfo{volume}{790} (\bibinfo{year}{2007})
  \bibinfo{pages}{10--16}.
\bibitem[{Polyakov(1999)}]{Polyakov:1998ze}
\bibinfo{author}{M.~V. Polyakov}, \bibinfo{journal}{Nucl. Phys. B}
  \bibinfo{volume}{555} (\bibinfo{year}{1999}) \bibinfo{pages}{231}.
\bibitem[{Polyakov and Weiss(1999)}]{Polyakov:1999gs}
\bibinfo{author}{M.~V. Polyakov}, \bibinfo{author}{C.~Weiss},
  \bibinfo{journal}{Phys. Rev. D} \bibinfo{volume}{60} (\bibinfo{year}{1999})
  \bibinfo{pages}{114017}.
\bibitem[{Accardi et~al.(l ex)}]{Accardi:2012qut}
\bibinfo{author}{A.~Accardi}, et~al.  (\bibinfo{year}{arXiv:1212.1701
  [nucl-ex]}). \bibinfo{note}{{\emph{Electron Ion Collider: The Next QCD
  Frontier - Understanding the glue that binds us all}}}.

\end{thebibliography}

\end{document}